\documentclass[twoside,fleqn]{article}

\usepackage{espcrc2,axodraw,graphics,epsfig}

\newcommand{\bea}{\begin{eqnarray}}
\newcommand{\eea}{\end{eqnarray}}
\newcommand{\beq}{\begin{equation}}
\newcommand{\eeq}{\end{equation}}

\newcommand{\pdir}{p\kern -5.2pt\raise 0.2ex\hbox {/}}
\newcommand{\vdir}{v\kern -5.75pt\raise 0.15ex\hbox {/}}
\newcommand{\kdir}{k\kern -5.75pt\raise 0.15ex\hbox {/}}
\newcommand{\epsdir}{\epsilon\kern -5.0pt\raise 0.15ex\hbox {/}}
\newcommand{\bvdir}{\bar{v}\kern -5.75pt\raise 0.15ex\hbox {/}}
\newcommand{\Ddir}{D\kern -7.75pt\raise 0.20ex\hbox {/}}
\newcommand{\Adir}{A\kern -7.75pt\raise 0.20ex\hbox {/}}
\newcommand{\ldir}{l\kern -5.0pt\raise 0.2ex\hbox{/}}
\newcommand{\varepsdir}{\varepsilon\kern -5.5pt\raise 0.15ex\hbox{/}}

\newcommand{\beqns}{\begin{eqnarray*}}
\newcommand{\eeqns}{\end{eqnarray*}}

\newcommand{\gc}{\gamma^5}

\newcommand{\lgl}{\langle}
\newcommand{\rgl}{\rangle}

\newcommand{\m}{\mu}

\newcommand{\msb}{\overline{\rm{MS}}}

\newcommand{\bsb}{\overline{B_s}}

\def\elematrice#1#2#3{\lgl#1|#2|#3\rgl}

\def\Journal#1#2#3#4{{#1} {\bf #2}, #3 (#4)}

\def\NPB{{\em Nucl. Phys.} B}

\def\PLB{{\em Phys. Lett.}  B}

\def\PRL{\em Phys. Rev. Lett.}
\def\PRD{{\em Phys. Rev.} D}

 \title{Lattice measurement of $B_{B_s}$ with a chiral light quark action}

\author{B. Blossier\address{DESY, 
        Platanenallee 6, D-15738 Zeuthen, Germany}}

\begin{document}

 \begin{abstract}
\noindent The computation on the lattice of the bag parameter $B_{B_s}$ associated to
the $B_s - \bsb$ mixing amplitude in the Standard Model is presented. The
estimation has been made by combining the static limit of HQET and the Neuberger
light quark action which preserves the chiral symmetry on the lattice. We find $B^{\msb\,
{\rm stat}}_{B_s}(m_b)=0.92(3)$.
\end{abstract}

\maketitle

\section{Introduction}
\noindent $B_s-\bsb$ mixing is highly important in testing the Standard 
Model (SM) and constrains strongly its extensions by bounding the unitarity triangle. Since it is a 
flavor changing 
neutral process it occurs through loops so that the corresponding mixing amplitude 
is a sensitive
 measure of $|V_{ts}|$ and $|V_{tb}|$, as the major SM loop contribution comes from 
 $t$-quark. The mixing of weak interaction eigenstates $B_s$ and 
$\bsb$ induces a mass gap $\Delta M_s$ between the mass eigenstates 
$B_{sH}$ and $B_{sL}$. Experimentally, D0 has bounded $\Delta M_s$ as 
$17 \; \rm{ps}^{-1} < \Delta M_s < 21 \; \rm{ps}^{-1} \; (90\, $\% CL) \cite{D0Bsbar}
and the measurement made at CDF gave $\Delta M_s = 17.330^{+0.426}_{-0.221}$ 
\cite{CDFBsbar}.

\noindent Theoretically the $B_s-\bsb$ mixing is described by means of an Operator 
Product Expansion, $i.e.$ the Standard Model Lagrangian ${\cal L}_{SM}$ is reduced to an 
effective Hamiltonian ${\cal H}_{eff}^{\Delta B=2}$, up to negligible terms of 
${\cal O}(1/M^2_W)$:
\bea\nonumber
{\cal H}^{\Delta B=2}_{\rm eff}&=&\frac{G^2_F}{16\pi^2}M^2_W 
(V^*_{tb}V_{ts})^2\\
&\times& \eta_B S_0(x_t) C(\mu_b)Q^{\Delta B=2}_{LL}(\mu_b),
\eea 
\bea\nonumber
Q^{\Delta B=2}_{LL}=\bar{b}\gamma_{\mu L}s\; \bar{b}\gamma_{\mu L}s, \quad 
\mu_b\sim m_b,
\eea
where $\eta_B=0.55\pm 0.01$. $S_0(x_t)$ is a known Inami-Lim function of $x_t=m^2_t/M^2_W$
\cite{inami}, 
$C(\mu_b)$ is the Wilson coefficient computed perturbatively to NLO in $\alpha_s(\mu_b)$ in 
the $\msb$ (NDR) scheme, and $Q^{\Delta B=2}_{LL}$ is a 
four-fermions operator coming from the reduction of the box diagrams in 
${\cal L}_{SM}$ to a local operator in the effective theory. The mass splitting is
\beq
\Delta M_{B_s} = 2 | \elematrice{\bsb}{{\cal H}^{\Delta B=2}_{\rm eff}}{B} |\; .
\eeq
The hadronic matrix element of $Q^{\Delta B=2}_{LL}$, which must be computed non perturbatively, 
is conventionally parameterized as
\beq
\lgl\overline{B_s}|Q^{\Delta B=2}_{LL}(\mu_b)|B_s\rgl \equiv \frac 8 3 m^2_{B_s} 
f^2_{B_s}B_{B_s}(\mu_b)\; ,
\eeq
where $B_{B_s}(\mu_b)$ is the $B_s$ meson bag parameter and $f_{B_s}$ the decay constant. In the
following it will be useful to introduce in addition to the operator 
$O_1 \equiv Q^{\Delta B=2}_{LL}$ the operators  of the so called supersymmetric basis
\bea\nonumber
O_2&=&\bar{b}^i (1-\gc) s^i \bar{b}^j (1-\gc) s^j,\\
\nonumber
O_3&=&\bar{b}^i\gamma_\m (1-\gc) s^i \bar{b}^j\gamma_\mu (1+\gc) s^j,\\
O_4&=&\bar{b}^i(1-\gc) s^i \bar{b}^j(1+\gc) s^j,
\eea
whose the matrix elements $\elematrice{\bsb}{O_i}{B_s}$ are parameterised by
\bea\nonumber
\elematrice{\bsb}{O_2}{B_s}&=&-\frac{5}{3}\left(\frac{m_{B_s}}{m_b(\m)+m_s(\m)}
\right)^2f^2_{B_s} B_2(\m),\\
\nonumber
\elematrice{\bsb}{O_3}{B_s}&=&-\frac{4}{3}\left(\frac{m_{B_s}}{m_b(\m)+m_s(\m)}
\right)^2f^2_{B_s} B_3(\m),\\
\elematrice{\bsb}{O_4}{B_s}&=&2\left(\frac{m_{B_s}}{m_b(\m)+m_s(\m)}
\right)^2f^2_{B_s} B_4(\m).
\eea
\noindent So far $B_{B_s}(\mu_b)$ has been computed by using lattice QCD 
\cite{bernard}-\cite{aoki2}\footnote{It was also estimated by using QCD sum rules 
\cite{narison} but we will concentrate only on lattice results.}. One of the major problems 
with those computations is in the 
following: the standard Wilson light quark lattice action breaks explicitely the chiral 
symmetry, which tremendously complicates the renormalization procedure of 
$Q^{\Delta B =2}_{LL}$ and its matching to the continuum. To get around that problem we 
compute $B_{B_s}(\mu_b)$ by using the lattice 
formulation of QCD in which the chiral symmetry is preserved at finite lattice spacing 
\cite{overlap}. On the other hand, it should be stressed that our heavy quark is static, 
as the currently available lattices do not allow to work directly with the propagating $b$ 
quark. Thus our results will suffer from $1/m_b$-corrections.

\section{Computation on the lattice}

\begin{figure}[htb]
\begin{center}
\vspace{-0.7cm}
\includegraphics*[width=5cm, height=3.8cm]{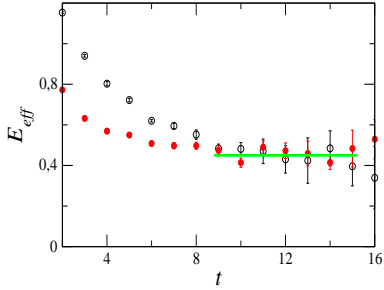}
\end{center}
\vspace{-0.8cm}
\caption{Effective binding energy of the $0^-$-state when currents are local (unfilled symbols) 
or smeared (filled symbols)}
\label{fig:smearing}
\end{figure}

\noindent In our numerical simulation we choose to work with the action 
$S=S^{\rm EH}_h+S^{\rm N}_l$, where 
\bea\nonumber
\hspace{-0.5cm}
S^{\rm EH}_h=a^3\sum_{x} 
\Big\{\bar{h}^+(x)\left[h^+(x)-V^{{\rm HYP}\dag}_0(x-\hat{0})h^+(x-\hat{0})\right]
\eea
\vspace{-0.5cm}
\bea\nonumber
\hspace{1.5cm}-\bar{h}^-(x)\left[V^{{\rm HYP}}_0(x)h^-(x+\hat{0})-h^-(x)\right]\Big\}
\eea 
is the static limit of HQET action \cite{eichten} which has been modified after using the so-called 
HYP (hypercubic blocking) procedure \cite{hyp}, that is enough to substantially improve the 
signal/noise ratio \cite{Dellamorte} [the field $h^+(h^-)$ annihilates the static heavy quark
(antiquark)]. 
$S^{\rm N}_l=a^3 \sum_x \bar{\psi}(x) D^{(m_0)}_N\psi(x)$ is the overlap light quark action with
\bea\nonumber
D^{(m_0)}_N=\left(1-\frac{1}{2\rho} am_0 \right)D_N+m_0,
\eea
\bea\nonumber 
D_N=\frac{\rho}{a}\left(1+\frac{X}{\sqrt{X^\dag X}}\right), \quad X=D_W -\frac{\rho}{a},
\eea
where $D_W$ is the standard Wilson-Dirac operator. The overlap Dirac operator $D^{(m_0)}_N$
verifies the Ginsparg-Wilson relation 
$\{\gamma^5,D^{(m_0)}_N\}=\frac{a}{\rho}D^{(m_0)}_N\gamma^5D^{(m_0)}_N$ and 
the overlap action is invariant under the chiral light quark transformation \cite{luscher}
\bea\nonumber
\psi \to \psi+i\epsilon \gamma^5\left(1-\frac{a}{\rho}D^{(m_0)}_N\right)\psi, \quad 
\bar{\psi}\to \bar{\psi}(1+i\epsilon \gamma^5),
\eea
which is essential to prevent mixing of four-fermion operators of different chirality 
\cite{becirevic}. In other words, in the renormalization 
procedure, the subtraction of the spurious mixing with $d=6$ operators will not be needed.
\noindent We thus compute the two- and three-point functions:
\beq\label{c2}
\tilde{C}^{(2)\pm}_{AA}(t)=\lgl\sum_{\vec{x}}\widetilde{A}^\pm_0(\vec{x},t)
\widetilde{A}^{\pm\dag}_0(0)\rgl_{_{U}}
\stackrel{t\gg 0}{\longrightarrow} \tilde{Z}_A e^{-\epsilon t},\\
\eeq
\vspace{-0.3cm}
\bea\nonumber\label{c3}
\tilde{C}^{(3)}_{VV+AA}(t_i,t)=\lgl\sum_{\vec{x},\vec{y}} 
\widetilde{A}^+_0(\vec{x},t_i)\widetilde{O}_1(0,0)
\widetilde{A}^{-\dag}_0(\vec{y},t)\rgl_{_{U}}
\eea
\vspace{-0.3cm}
\bea
\hspace{1.4cm}\stackrel{t_i-t \gg 0}{\longrightarrow}\tilde{Z}_A\, _v\lgl
\overline{B_s}|\widetilde{O}_1(\mu)|B_s\rgl_v\; e^{-\epsilon(t_i-t)},
\eea
\vspace{-0.3cm}
\bea\nonumber\label{c32}
\tilde{C}^{(3)}_{SS+PP}(t_i,t)=\lgl\sum_{\vec{x},\vec{y}} 
\widetilde{A}^+_0(\vec{x},t_i)\widetilde{O}_2(0,0)
\widetilde{A}^{-\dag}_0(\vec{y},t)\rgl_{_{U}}
\eea
\vspace{-0.3cm}
\bea\hspace{1.4cm}
\stackrel{t_i-t \gg 0}{\longrightarrow} \tilde{Z}_A\,\, _v\lgl \overline{B_s}|
\widetilde{O}_2(\mu)|B_s\rgl_v\; e^{-\epsilon(t_i-t)},
\eea
\bea\nonumber
\widetilde{A}^\pm_0\equiv \bar{h}^\pm\gamma_0\gamma^5 s, \quad 
\widetilde{O}_1=\bar{h}^{(+)i}\gamma_{\mu L}s^i\bar{h}^{(-)j}
\gamma_{\mu L}s^j,
\eea
\vspace{-0.7cm}
\bea\nonumber
\widetilde{O}_2=\bar{h}^{(+)i}P_L s^i\bar{h}^{(-)j}P_L s^j,
\sqrt{\tilde{Z}_A}=\lgl 0|\widetilde{A}^\pm_0|B_s\rgl_v.
\eea
$\epsilon$ is the binding energy of the 
pseudoscalar heavy-light meson. In the computation of $\tilde{C}^{(2)\pm}(t_i,t)$ one 
current $\tilde{A}^\pm_0$ 
is local whereas the other is smeared. 
The role of the smearing is to isolate earlier the ground state \cite{boyle}, as shown in Fig.
\ref{fig:smearing}~\footnote{Even if the time interval from which we extract the binding energy 
starts at $t=9$ (green line), the overlap with radial excitations is
quite reduced since $t=6$ when currents are smeared.}. We see that the same state is isolated when 
purely local currents are used (with those currents the signal does not exist if 
$V^{\rm HYP}_0$ is not used in the heavy quark action).
The source operators in $\tilde{C}^{(3)}_{VV+AA}(t_i,t)$ and $\tilde{C}^{(3)}_{SS+PP}(t_i,t)$ 
are the smeared currents $\tilde{A}^\pm_0$, whereas the four-fermion operators $\tilde{O}_1$ 
and $\tilde{O}_2$ are purely local.
In (\ref{c2}), (\ref{c3}) and (\ref{c32}) the subscript "$v$" and superscript "$\sim$" are 
designed to remind the reader that states and operators are defined in HQET. 

\noindent Note that in the 
computation of $\tilde{C}^{(3)}_{VV+AA}(t_i,t)$ and $\tilde{C}^{(3)}_{SS+PP}(t_i,t)$ there
are two terms, coming from 
two different Wick contractions: 
\bea\nonumber 
\sum_i B_{ii}(t)\displaystyle{\sum_j}B_{jj}
(t_i)\quad {\rm and} \quad \sum_{i,j}B_{ij}(t)B_{ji}(t_i). 
\eea
$i$, $j$ are the color indices and
\bea\nonumber
B_{ij}(t)={\rm Tr}\left[\sum_{\vec{x}} \gamma_{\mu L}{\cal S}^{\dag ik}_L(0;\vec{x},t) \gamma_0\gamma^5
{\cal S}^{kj}_H(\vec{x},t;0)\right];
\eea 
${\cal S}_L$ and ${\cal S}_H$ are the light and heavy propagators 
respectively and the trace is over spinor indices.

\begin{figure}[htb]
\begin{center}
\vspace{-0.7cm}
\includegraphics*[width=5cm, height=3.8cm]{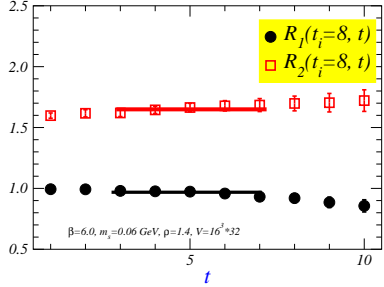}
\end{center}
\vspace{-0.8cm}
\caption{Signals for $R_{1,2}(t_i,t)$ defined in eq (\ref{rap}): lines
indicate the time interval on which we fit the signal to a constant to extract
$\tilde{B}_1(a)$ and $\tilde{B}_2(a)$ respectively.}
\label{fig:rap}
\end{figure}

\noindent After having computed the correlation functions (\ref{c2}), (\ref{c3}) and
(\ref{c32}) we build the following two ratios $R_1(t_i,t)$ and $R_2(t_i,t)$:
\bea\label{rap}
\nonumber
R_1(t_i,t)&=&\frac{\tilde{C}^{(3)}_{VV+AA}(t_i,t)}{\frac 8 3 \tilde{Z}^2_A \tilde{C}^{(2)+}_{AA}(t_i)
\tilde{C}^{(2)-}_{AA}(t)}\\
\nonumber
&\stackrel{t_i-t \gg 0}{\longrightarrow}&
\frac{_v\lgl \overline{B_s}|\tilde{O}_1|B_s\rgl_v}{\frac 8 3 |\lgl 0|\tilde{A}^-_0|B_s\rgl_v|^2}
\equiv \tilde{B}_1(a),\\
\nonumber
R_2(t_i,t)&=&\frac{C^{(3)}_{SS+PP}(t_i,t)}{-\frac 5 3 \tilde{Z}^2_A \tilde{C}^{(2)+}_{AA}(t_i)
\tilde{C}^{(2)-}_{AA}(t)}\\
&\stackrel{t_i-t \gg 0}{\longrightarrow}&
\frac{_v\lgl \overline{B_s}|\tilde{O}_2|B_s\rgl_v}{-\frac 5 3 |\lgl 0|\tilde{A}^-_0|B_s\rgl_v|^2}
\equiv \tilde{B}_2(a).
\eea
Those ratios are calculated either with a fixed time $t \in [-6,-8,-10,-12,-14,-16]$ and  
$t_i$ free, or by fixing $t_i \in [6,8,10,12,14,16]$ while letting $t$ free. We take the 
average of the two options. In Fig. \ref{fig:rap} we show the quality of the signals for
$R_{1,2}(t_i,t)$, with $t_i=6$ fixed. The signal for $\tilde{B}_1(a)$ is quite
stable as a function of $t_i$, whereas the signal for $\tilde{B}_2(a)$ rapidly deteriorates for 
larger $t_i$, and is completely lost for $t_i>10$.

\section{Extraction of physical $B_{B_s}$}

\noindent Three steps are required to extract $B_{B_s}\equiv B_1$ from the lattice:\\
(1) $\tilde{B}_{1,2}(a)$ are matched onto the continuum $\overline{\rm MS}$(NDR) scheme at NLO in
perturbation theory at the renormalization scale $\mu=1/a$ \cite{becirevic},\\
(2) $\tilde{B}_{1,2}$ are evolved from $\mu=1/a$ to $\mu=m_b$ by using the HQET anomalous
dimension matrix, known to 2-loop accuracy in perturbation theory \cite{spqcdr,run},\\
(3) $\tilde{B}_{1,2}(\mu=m_b)$ are then matched onto their QCD counterpart, $B_{1,2}(m_b)$, 
in the $\overline{\rm MS}$(NDR) scheme at NLO \cite{run}.\\
The scales chosen to do the matchings are such that neither $\ln (a\mu)$ in the step (1) nor 
$\ln(\mu/m_b)$ in the step (3) correct strongly the matching constants. The advantage 
of using a chiral light 
quark action for the step (1) lies in the fact that four-fermion operators can mix only with 
a dimension 6 four-fermion operator of the same chirality. In other words we have not more 
than 4 independent renormalization 
constants in the renormalization matrix, because $\tilde{O}_1$ and $\tilde{O}_2$ can mix 
neither with $\tilde{O}_3\equiv \bar{h}^+\gamma_{\mu L}s\;\bar{h}^-\gamma_{\mu R}s$, nor with 
$\tilde{O}_4\equiv \bar{h}^+(1-\gamma^5)s\;\bar{h}^-(1+\gamma^5)s$. 
Actually, thanks to the heavy quark symmetry, those constants are not all independent and we 
have \cite{becirevic}
\bea\nonumber
\left(\begin{array}{c}
\tilde{B}^{\overline{\rm MS}}_1(\mu)\\
\tilde{B}^{\overline{\rm MS}}_2(\mu)
\end{array}
\right)
=
\left(\begin{array}{cc}
Z_{11}(a\mu)&0\\
\frac{Z_{22}(a\mu)-Z_{11}(a\mu)}{4}&Z_{22}(a\mu)
\end{array}\right)
\left(\begin{array}{c}
\tilde{B}_1(a)\\
\tilde{B}_2(a)
\end{array}\right).
\eea
\section{Results and discussion}
\noindent Our results are based on two simulations, with the parameters given in Tab. 
\ref{table}.
\begin{table}
\begin{center}
\vspace{-0.5cm}
\begin{tabular}{|c|c|c|c|c|c|c|}
\hline
$\beta$&${\rm N_{conf}}$&action&$\rho$&$am^s_0$&$\kappa_l$\\
\hline
6.0&100&overlap&1.4&0.06&\\
&&Wilson&&&0.1435\\
\hline
5.85&40&overlap&1.6&0.09&\\
\hline
\end{tabular}
\end{center}
\caption{Parameters of our simulations: $am^s_0$ and $\rho$ have been chosen following
\cite{giusti,hernandez}; the volume is $16^3\times 32$.}
\label{table}
\end{table} 

\begin{figure}[t]
\begin{center}
\begin{picture}(185,30)(65,-10)

\LinAxis(0,40)(160,40)(3,4,-3.2,0,0.2)
\LinAxis(0,-140)(160,-140)(3,4,3.2,0,0.2)
\Line(0,40)(0,-140) \Line(160,40)(160,-140)
\DashLine(-20,-110)(160,-110){2}

\Text(0,-150)[]{0.7}
\Text(40,-150)[]{0.8}
\Text(80,-150)[]{0.9}
\Text(120,-150)[]{1.0}
\Text(160,-150)[]{1.1}

\Text(20,30)[]{\small{$0.97(12)$}}
\Text(20,10)[]{\small{$0.81(7)$}}
\Text(20,-10)[]{\small{$0.93(^{+08}_{-10})$}}
\Text(20,-30)[]{\small{$0.90(^{+4}_{-2})$}}
\Text(20,-50)[]{\small{0.87(2)}}
\Text(20,-70)[]{\small{0.85(5)}}
\Text(20,-90)[]{\small{$0.85(6)$}}
\Text(20,-120)[]{\small{$0.92(3)$}}


\Line(75,-120)(105,-120)
\Line(75,-122)(75,-118)
\Line(105,-122)(105,-118)
\CTri(87,-120)(90,-116)(90,-124){Blue}{Blue}
\CTri(90,-116)(90,-124)(93,-120){Blue}{Blue}

\Line(38,-100)(81,-100)
\Line(38,-102)(38,-98)
\Line(81,-102)(81,-98)
\CBoxc(60,-100)(4,4){Red}{Red}

\Line(40,-80)(80,-80)
\Line(40,-82)(40,-78)
\Line(80,-82)(80,-78)
\CBoxc(60,-80)(4,4){Red}{Red}

\Line(60,-60)(80,-60)
\Line(60,-62)(60,-58)
\Line(80,-62)(80,-58)
\CTri(67,-63)(70,-56)(73,-63){Red}{Red}


\SetColor{Black}
\Line(72,-40)(95,-40)
\Line(72,-42)(72,-38)
\Line(95,-42)(95,-38)
\CCirc(80,-40){3}{Red}{Red}

\SetColor{Black}
\Line(52,-20)(122,-20)
\Line(52,-22)(52,-18)
\Line(122,-22)(122,-18)
\CCirc(92,-20){3}{Red}{Red}

\SetColor{Black}
\Line(12,0)(72,0)
\Line(12,-2)(12,2)
\Line(72,-2)(72,2)
\CCirc(42,0){3}{Blue}{Blue}

\SetColor{Black}
\Line(70,20)(158,20)
\Line(70,18)(70,22)
\Line(158,18)(158,22)
\CCirc(114,20){3}{Red}{Red}

\Text(168,-120)[l]{\tiny{Orsay, static heavy quark (2006)}}

\Text(168,-100)[l]{\tiny{JLQCD (unq. $N_f=2$), NRQCD (2003)}}

\Text(168,-80)[l]{\tiny{JLQCD, NRQCD (2003)}}

\Text(168,-60)[l]{\tiny{SPQcdR (2002)}}

\Text(168,-40)[l]{\tiny{UKQCD (2001)}}

\Text(168,-20)[l]{\tiny{APE (2000)}}

\Text(168,0)[l]{\tiny{Gimenez and Reyes (1999)}}

\Text(168,20)[l]{\tiny{Bernard, Blum and Soni (1998)}}

\rText(-10,-30)[][l]{\small{Renormalisation with subtractions}}
\rText(-14,-125)[][l]{\small{without}}
\rText(-7,-125)[][l]{\small{subtr.}}

\end{picture}
\end{center}
\vspace{4.6cm}
\caption{ \label{fig:comp} Various lattice values of $B^{\overline{MS}}_{B_s}(m_b)$
\cite{bernard}-\cite{aoki2}; blue symbols correspond to a computation made with a static heavy 
quark}
\end{figure}
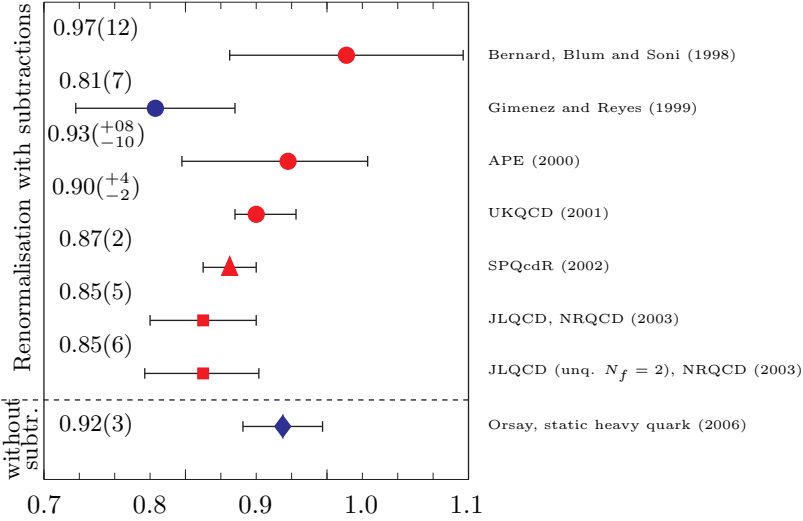

\begin{table}
\begin{center}
\begin{tabular}{ll}
\begin{tabular}{|c|c|}
\hline
$\frac{\lgl\widetilde{O_3}\rgl^{\rm lat}}
{\lgl\widetilde{O_1}\rgl^{\rm lat}}$
&-1.011(1)\\ 
\hline
$\frac{\lgl\widetilde{O_4}\rgl^{\rm lat}}
{\lgl\widetilde{O_1}\rgl^{\rm lat}}$
&1.013(2)\\
\hline
\end{tabular}
&
\begin{tabular}{|c|c|c|c|}
\hline
HYP&$Z_{13}$&$Z_{14}$&$B^{\rm stat}_{B_s}(m_b)$\\
\hline
no&-0.459&-0.919&0.763(5)\\
yes&-0.235&-0.470&0.873(5)\\
\hline
\end{tabular}\\
\end{tabular} 
\end{center}
\caption{Ratios $\lgl\widetilde{O}_{3,4}\rgl^{\rm lat}/\lgl\widetilde{O}_1\rgl^{\rm lat}$ extracted from the
simulation with a Wilson light quark and the improved HQET action (left table); comparison of 
$Z_{13}$, $Z_{14}$ and  $B^{\rm stat}_{B_s}(m_b)$ obtained with the Wilson action in function of HQET action 
improvement (right table). \label{tab:ren}}
\end{table}

\noindent We find $B^{\overline{\rm MS}\, {\rm stat}}_{B_s}(m_b)=0.922(12)(25)$, where the first error is statistical,
the second is systematic and contains the error from the estimation of $\alpha_s(1/a)$ and the 
finite $a$ effects. From Fig. \ref{fig:comp} 
it can be seen that our value is larger than the previous static
result \cite{gimenez}. This difference is likely due to the use of 
Neuberger light quark action (no subtractions), due to the use of the HYP procedure, or
the combination of both. To answer to this question we made a computation with the Wilson light
quark action. In that case we have subtractions in the renormalisation procedure:
\bea\nonumber
\lgl\widetilde{O}_1\rgl^{\rm con}= Z_{11} \lgl\widetilde{O}_1\rgl^{\rm lat} \left(1+Z_{13}
\frac{\lgl\widetilde{O}_3\rgl^{\rm lat}}{\lgl\widetilde{O}_1\rgl^{\rm lat}}
+Z_{14}\frac{\lgl\widetilde{O}_4\rgl^{\rm lat}}{\lgl\widetilde{O}_1\rgl^{\rm lat}}\right).
\eea 
A correction on $Z_{13}$ and $Z_{14}$, coming
from a different definition of the HQET action,
induces a substantial systematic error on $B_{B_s}$ as illustrated on the Tab. \ref{tab:ren}.
From Fig. \ref{fig:comp} we also notice that our value is also somewhat 
larger than the results obtained with the 
propagating heavy quark. We can take account of the $1/m_b$ effects by interpolating linearly through
$M_{B_s}$:
\beq
B^{\msb, M_{B_s}}_{B_s} (m_b)=B^{\msb, {\rm stat}}_{B_s} (m_b)\left(1+\frac{C}{M_{B_s}}\right),
\eeq
where $C=-0.24(6)$ GeV \cite{ape}. With $M_{B_s}=5.37$ GeV, 
$B^{\msb, M_{B_s}}_{B_s}=0.955(11)\times B^{\msb, {\rm stat}}_{B_s} (m_b)$, we obtain after the simulation
at $\beta=6.0$ $B^{\msb, M_{B_s}}_{B_s}=0.881(15)$.
JLQCD collaboration showed that the errors due to 
quenching are likely to be small \cite{aoki1,aoki2}. That issue has to be addressed  by unquenching 
the $B_s-\bsb$ mixing amplitude in the static limit and by avoiding the subtraction procedure as well. 
The first step would be to make a simulation with 2 degenerate Wilson sea light quarks and an overlap
valence strange quark.

\noindent Eventually with $f^{\rm unq}_{B_s}=230$ MeV \cite{hashimoto}, $V_{ts}=04076$ \cite{utfit}, 
$V_{tb}=0.99912$
\cite{ckmfit}, we obtain $\Delta M^{\rm SM}_{B_s}=20.7\, {\rm ps}^{-1}$. However this value has to be taken
very carefully because the uncertainty on $f^{\rm unq}_{B_s}$ is 30\%.

\section{Conclusion} \noindent For the first time the bag parameter associated to the $B_s - \bsb$ mixing 
amplitude in the Standard Model has been computed on the lattice by combining the static
limit of HQET and a light quark action 
which preserves
the chiral symmetry on the lattice. Thus systematic error induced by subtractions in the 
renormalisation
procedure are absent, since there is no mixing among dimension 6 four-fermion operators of 
different
chirality. $1/m_b$ corrections and quenching effects have still to be studied carefully.

\end{document}